\documentclass[seceq]{ptptex}

\usepackage{enumerate}
\usepackage{graphicx}

\newcommand{\be}{\begin{equation}}
\newcommand{\ee}{\end{equation}}
\newcommand{\bea}{\begin{eqnarray}}
\newcommand{\eea}{\end{eqnarray}}
\newcommand{\f}{\frac}  
\newcommand{\la}{\lambda}

\newcommand{\da}{\downarrow}

\newcommand{\Ga}{\Gamma}
\newcommand{\ga}{\gamma}

\newcommand{\ket}{\rangle}

\newcommand{\av}{{\rm{av}}}
\newcommand{\fl}{{\rm{fl}}}

\newcommand{\ov}{\overline}

\newcommand{\De}{\Delta}
\title{%        %You can use \\ for explicit line-break
Energy averages over regular and chaotic 
states in the decay out of superdeformed bands%
}

\author{%       %Use \scshape  for the family name
Mahir S. \textsc{Hussein}$^1$,
Adam J. \textsc{Sargeant}$^1$,
Maur\'\i cio P. \textsc{Pato}$^1$,
Noboru \textsc{Takigawa}$^2$
and Manabu \textsc{Ueda}$^3$}

\inst{%         %Affiliation, neglected when [addenda] or [errata]
{$^1$Instituto de F\'\i sica, Universidade de S\~{a}o Paulo,
Caixa Postal 66318, 05315-970 S\~{a}o Paulo, SP, Brazil}
\\$^2$Department of Physics, Tohoku University, Sendai, 980-8578, Japan
\\$^3$Akita National College of Technology, 
Iijima Bunkyo-cho 1-1, Akita, 011-8511, Japan}

\abst{%         %this abstract is neglected when [addenda] or [errata]
We describe the decay out of a superdeformed band using the methods of 
reaction theory. Assuming that decay-out occurs due to equal coupling
(on average) to a sea of equivalent chaotic normally deformed (ND) states,
we calculate the average intraband decay intensity 
and show that it can be written as an ``optical'' background term plus a
fluctuation term, in total analogy with average nuclear cross sections.
We also calculate the variance in closed form. We investigate how these 
objects are modified when the decay to the ND states occurs via 
an ND doorway \emph{and} the ND states' statistical properties are changed from
chaotic to regular. We show that the average decay intensity depends on 
two dimensionless variables in the first case  
while in the second case, four variables enter the picture.
}

\begin{document}

\maketitle

\section{Introduction}

The first superdeformed (SD) rotational band to be observed 
was that identified in the nucleus $^{152}_{66}$Dy$_{86}$ by 
Twin \emph{et al.}\cite{Twin:1986}. 
A sequence of nineteen $\ga$-rays of nearly
constant spacing was observed which permitted the attribution of the
moment of inertia of a symmetric prolate rigid rotor with elliptic axes
in the proportion 1:1:2. Since then 320 SD bands have been observed
in various
mass regions extending from $A$$\sim$20 to $A$$\sim$240\cite{Singh:2002}.
The stability of nuclei at any deformation can be related to the existence
of energy gaps between shells of independent particle states. 
The shell gaps
which appear for certain nuclear deformations give origin to non-spherical
configurations of special stability\cite{Nilsson:1995,Wadsworth:2002}.

As explained by Lopez-Martens \emph{et al.}\cite{Lopezmartens:2003},
the $\ga$-spectrum of a compound nucleus whose 
decay path includes an SD band typically contains $\ga$-rays
which result from four distinct stages:
\begin{enumerate}[I]
\item Statistical $\ga$-rays which populate the SD band after formation of
  the nucleus under consideration by particle evaporation following a 
  fusion evaporation or fragmentation reaction.
\item Collective $E$2 $\ga$-rays from decay along the SD band.
\item Statistical $\ga$-rays from excited normally deformed
  (ND) configurations.
\item Collective $E$2 $\ga$-rays from decay along the yrast ND band.
\end{enumerate}
In cases where this kind of decay scheme applies, stages I and III
correspond to the cooling of a hot and chaotic system while stages
II and IV correspond to the decay of a cold and regular system. Other
decay schemes which include SD bands are possible. An SD
state may fission or emit particles instead of decaying to the
ground state by $\ga$-emission. Stage II shows the following intensity
profile as a function of transition energy, rotational frequency or 
spin:\cite{Lopezmartens:2000}
gradual population at high spin, a plateau at intermediate spin and
sudden depopulation at low spin (between 10 and 30 $\hbar$ depending
on the mass region).

SD bands by definition are found in the second minimum of the
potential energy surface in deformation space (hyperdeformed bands in
the third if they can be observed). For a certain interval in spin
this minimum coexists with the first minimum where the ND states are
found. Between these two minima there is a barrier of typically a few
MeV in height. Rotational bands exist in both minima over a large
interval in spin. For low spins an ND band is yrast while for higher
spins an SD band is yrast. At a certain spin these two bands cross.
Decreasing in spin from the crossing point the excitation energy of
the SD band relative to the ND band increases due to the different 
moments of inertia of the two bands. At some moment states in the SD 
band begin to decay to states in the ND minimum. This transition 
between stages II and III proceeds by tunneling through
the multidimensional barrier in deformation 
space\cite{Shimizu:1992,Yoshida:2001}.
 
In most cases it has not been possible to connect SD bands to the rest
of the known decay scheme and consequently only $\ga$-transition
energies, lifetimes and branching rations are known. The reason is the
complexity of the ND decay scheme and our incomplete knowledge of it.
In just 10\% of the 320 bands identified have absolute excitation
energies, spins and parities been
designated\cite{Lauritsen:2002,Wilson:2003,Paul:2001}.

In the present paper we are interested in stage II of the 
$\ga$-intensity. In fact we shall not discuss the
feeding\cite{Khoo:1993} of SD
bands and restrict ourselves to the decay out. 
In Section \ref{GOE} we present analytic formulae for 
the energy average 
(including the energy average of the fluctuation contribution) and 
variance of the intraband decay intensity of a SD band
in terms of variables which usefully describe the decay-out
\cite{Krucken:2001we,Gu:1999bv,Sargeant:2002sv}. 
In agreement with Gu and Weidenm\"uller \cite{Gu:1999bv} (GW)
we find that average of the total intraband decay
intensity can be written as a function of the dimensionless variables
$\Ga^\da/\Ga_S$ and $\Ga_N/D$ where
$\Ga^\da$ is the spreading width for the 
mixing of an SD state with ND states of the
same spin, $D$ is the mean level spacing of the latter and 
$\Ga_S$ $(\Ga_N)$ are the electromagnetic decay widths of the SD (ND) 
states. Our formula for the variance of the total intraband decay 
intensity,
in addition to the two dimensionless variables just mentioned,
depends on the dimensionless variable $\Ga_N/(\Ga_S+\Ga^\da)$. 

The results of Section~\ref{GOE} depend on two statistical assumptions.
Firstly it is assumed that the states in the ND 
minimum are chaotic and that their statistical properties can be 
described by the GOE. Secondly, it is assumed that on average the 
SD band couples with equal strength to all ND states. 
However it might be appropriate in some mass
regions where the decay out occurs at lower excitation energy 
to describe the ND states by Poisson statistics. 
Further, it may be incorrect to assume that all ND states are 
equivalent. Certain ND states which have a stronger overlap with the
SD states may serve as doorways\cite{Andreoiu:2003} 
to the remaining ND configurations. 
In Section~\ref{Poisson} a model for which the ND states obey Poisson
statistics and for which an SD state couples to a single ND doorway
is discussed\cite{Sargeant:2001aq,Aberg:1999a}. It is shown that the
addition variables become important.

\section{Energy average and variance of the decay intensity}
\label{GOE}
The total intraband decay intensity has the form 
\cite{Sargeant:2001aq,Gu:1999bv}
\be
I=\left(2\pi\Ga_S\right)^{-1}\int_{-\infty}^{\infty}
dE|A_{00}(E)|^2,
\label{I}
\ee
where $A_{00}(E)$ is the intraband decay amplitude 
and $\Ga_S$ is the electromagnetic decay width of 
SD state $|0\ket$.
The energy average of Eq.~(\ref{I}) may be written 
as the incoherent sum \cite{Sargeant:2002sv,Kawai:1973}
$\ov{I}=I_\av+\ov{I_\fl}$,
where
\be
I_\av=\ov{I_\av}=\left(2\pi\Ga_S\right)^{-1}
\int_{-\infty}^{\infty}dE
\left|\ov{A_{00}(E)}\right|^2
\label{Iav}
\ee
and
\be
\ov{I_\fl}=\left(2\pi\Ga_S\right)^{-1}
\int_{-\infty}^{\infty}dE
\ov{\left|A_{00}^\fl(E)\right|^2},
\label{ovIfl}
\ee
where we have written the decay amplitude as
$A_{00}=\ov{A_{00}}+A_{00}^\fl$ where $\ov{A_{00}}$ is a background 
contribution and $A_{00}^\fl$ is the fluctuation on that background.
Sargeant \emph{et al.}\cite{Sargeant:2002sv} took the background to be
\be\label{2Aav}
\ov{A_{00}}=\f{\Ga_S}{E-E_0+i(\Ga_S+\Ga^\da)/2}.
\ee
Eq.~(\ref{2Aav}) exhibits the structure of an isolated
doorway resonance. The doorway $|0\ket$ has an escape width $\Ga_S$ for 
decay to the SD state with next lower spin and a spreading width 
$\Ga^\da=2\pi v^2/D$ where $v^2$ is the mean square coupling of $|0\ket$
to the ND states whose level spacing is $D$.

The auto-correlation function of the decay amplitude is 
given by\cite{Sargeant:2002sv}
\bea
\ov{A_{00}^\fl(E){A_{00}^\fl(E')}^*}
&\approx&2\left(2\pi\Ga_N/D\right)^{-1}
(\Ga^\da/\Ga_S)^2\hspace{2mm}
{\ov{A_{00}(E)}}^2\hspace{2mm}\f{i\Ga_N}{E-E'+i\Ga_N}\ov{{A_{00}(E')}^*}^2.
\label{2Aflauto}
\eea
When $E'=E$ this reduces to  
\be
\ov{\left|A_{00}^\fl\right|^2}=2\left(2\pi\Ga_N/D\right)^{-1}
\f{{\Ga_S}^2{\Ga^\da}^2}
{\left[(E-E_0)^2+(\Ga_S+\Ga^\da)^2/4\right]^2},
\label{Aflsqu}
\ee
which is the average of the fluctuation contribution to the transition 
intensity.

The integrals in Eqs.~(\ref{Iav}) and (\ref{ovIfl}) may be carried out
using the calculus of residues. One obtains 
\be\label{2Iav}
I_\av^{\rm{GOE}}=(1+\Ga^\da/\Ga_S)^{-1},
\ee
for the average background contribution and
\bea
\ov{I_\fl}&=&2\left(\pi\Ga_N/D\right)^{-1}
\f{\left(\Ga^\da/\Ga_S\right)^2}
{\left(1+\Ga^\da/\Ga_S\right)^3}
\label{2ovIfl}
=2\left(\pi\Ga_N/D\right)^{-1}
I_\av\left(1-I_\av\right)^2,
\label{3ovIfl}
\eea
for the average fluctuation contribution to the average decay intensity.
Eq.~(\ref{2ovIfl}) for $\ov{I_\fl}$ was compared to the
fit formula, obtained by GW,
\bea\nonumber
\ov{I_\fl^{\rm{GW}}}&=&\left[1-0.9139\left(\Ga_N/D\right)^{0.2172}
\right]
\exp\left\{-\f{\left[0.4343\ln\left(\f{\Ga^\da}{\Ga_S}\right)
-0.45\left(\f{\Ga_N}{d}\right)^{-0.1303}\right]^2}
{\left(\Ga_N/D\right)^{-0.1477}}\right\}.
\\\label{gufit}
\eea
and qualitative agreement of the two formulae
found\cite{Sargeant:2002sv}.  
The dependence of $\ov{I_\fl}$ (and that of $I_\av$) 
on $\Ga^\da/\Ga_S$ results from the resonant doorway energy 
dependence of the decay amplitude $\ov{A_{00}(E)}$ [Eq.~(\ref{2Aav})]. 
This energy dependence also manifests itself in the average of the fluctuation
contribution to the transition intensity $\ov{|A^\fl_{00}(E)|^2}$ 
[Eq.~(\ref{Aflsqu})]. GW include precisely the same energy dependence
in their calculation by use of an energy dependent transmission coefficient to 
describe decay to the SD band.
This is the reason for our qualitative agreement with
GW concerning $\ov{I}$. 

A measure of the dispersion of the calculated $I$ is given by the
variance $\ov{\left(\Delta I\right)^2}
=\ov{\left(I-\ov{I}\right)^2}$.
It was shown\cite{Sargeant:2002sv} that
\be\label{3var}
\ov{\left(\Delta I\right)^2}=
{\ov{I_\fl}}^2f_1\left(\xi\right)
+2I_\av\ov{I_\fl}f_2\left(\xi\right),
\ee
where the variable $\xi$ is defined by
\bea
\xi\equiv\f{\Ga_S+\Ga^\da}{\Ga_N}
=\f{\Ga_S}{\Ga_N}(1+\Ga^\da/\Ga_S)
=\f{\Ga_S}{D}\f{D}{\Ga_N}{I_\av}^{-1},
\label{xi}
\eea
and
\be
f_1\left(\xi\right)=\f{1}{\left(1+\xi\right)}+\f{\xi}{\left(1+\xi\right)^2}
+\f{\xi^2}{2\left(1+\xi\right)^3}
\hspace{.5cm}\mbox{and}\hspace{.5cm}
f_2\left(\xi\right)=\f{1}{2\left(1+\xi\right)}.
\ee

Since the variance depends only on $(\Ga_S+\Ga^\da)/\Ga_N$ in addition to
$\Ga^\da/\Ga_S$ and $\Ga_N/D$, upon fixing the latter two variables
the variance may be considered a function of any {\em one} of $\Ga^\da/\Ga_N$,
$\Ga_S/\Ga_N$, $\Ga^\da/D$ or $\Ga_S/D$ [see Eq.~(\ref{xi})]. 
Our result for the variance of the decay intensity, 
$\ov{\left(\Delta I\right)^2}$ [Eq.~(\ref{3var})] has a structure 
reminiscent of Ericson's expression for the variance of the cross section 
\cite{Ericson:1963}. 
In the case compound nucleus scattering, extraction of the correlation
width from a measurement of cross section autocorrelation function
permits the determination of the density of states of the compound nucleus
\cite{Ericson:1966}. In the present case the variance supplies 
a second ``equation'' besides that for $\ov{I}$.
Both equations are functions of $\Ga^\da$ and $D$, since the 
electromagnetic widths are measured. 
Thus both $\Ga^\da$ and $D$ can be unambiguously 
determined. The derivation in this section is 
strictly valid only for $\Ga_N/D\gg1$\cite{Sargeant:2002sv}. 
However the formulae were found to be  qualitatively correct even for 
$\Ga_N<1$\cite{Sargeant:2002sv}.

\section{Regularity versus chaos in the ND minimum}
\label{Poisson}

\AA berg\cite{Aberg:1999a} has suggested that an order-chaos 
transition in the ND states enhances the tunneling probability between
the SD and ND wells and consequently that the decay-out of SD bands is 
an example of ``chaos assisted tunneling''. 
Crucial to the argument\cite{Sargeant:2001aq} 
is a postulate that the decay-out occurs via 
an ND doorway state. An ND doorway state can be visualized as being
the tail in the ND minimum of the wavefunction of the zero-point 
vibration in the SD minimum which may be constructed by the 
Generator Coordinate Method (GCM) of 
Bonche \emph{et al.}\cite{Bonche:1990}. Microscopically the vibration
is a coherent superposition of 1p-1h states which is damped by the two
body residual interaction. Depending on the strength of the residual
interaction the 2p-2h states will be regular, chaotic or intermediate
between these two limits.
Decreasing in spin from the point where the SD and ND bands
cross, the excitation energy of the SD band relative to the ND band 
increases. Near the crossing point the residual interaction is weak, 
therefore the
ND states may be characterized by quantum numbers which are approximately 
conserved and their energy spectrum exhibits degeneracies. 
In the language of quantum chaos the ND states are regular and
obey Poisson statistics.
As the spin decreases the residual interaction grows with
the increasing excitation energy, destroying degeneracies and
increasing the complexity of the ND states until the regime of quantum 
chaos is reached as characterized by the GOE. 
This is a plausible picture of
the evolution of the nuclear Hamiltonian with decreasing spin. 
Given that it is correct, how is the cascade down the superdeformed band
modified by an order-chaos transition in the ND minimum?

\AA berg\cite{Aberg:1999a} constructed a                   
random matrix model for the ND states which interpolates between
Poisson and GOE statistics by varying the chaoticity  parameter $\la$
continuously in the range $0<\la\le 1$.  
The parameter $\la$ which is the determined by the
ratio of the variances of the diagonal and off-diagonal random Hamiltonian 
matrix elements simulates the effect of the residual interaction. 
The SD state is assumed to lie in middle of the ND spectrum and the 
doorway is chosen to be an ND state which lies halfway between the SD 
state and the edge of the ND spectrum. The tunneling probability is
given by $T_{\rm{SD}\rightarrow\rm{ND}}=\sum|a'_{\mu}|^2$  where 
$a'_{\mu}$ is the amplitude of the nuclear wavefunction in ND state
$\mu$. The broadening of the distribution of the $a'_{\mu}$ by the 
residual interaction results in an increase in 
$T_{\rm{SD}\rightarrow\rm{ND}}$ which is denoted
``chaos assisted tunneling''. The enhancement in the tunneling 
probability due to the transition from regularity to chaos is 
deduced from the ratio of $T_{\rm{SD}\rightarrow\rm{ND}}$ for 
$\la=0.001$ to $T_{\rm{SD}\rightarrow\rm{ND}}$ for $\la=0.1$. The
value of this ratio is estimated to be $N$ where $N$ is the number
of ND states. 
 
Although this idea is also plausible it is not clear what the
effect of the change in $T_{\rm{SD}\rightarrow\rm{ND}}$ 
calculated by \AA berg\cite{Aberg:1999a} has on the average of 
the total relative decay intensity of the collective $E$2 electromagnetic 
transitions of SD bands where the sudden attenuation is 
experimentally observed. We calculated\cite{Sargeant:2001aq} 
the background contribution to
this quantity for \AA berg's model using the reaction theory 
methods which were outlined in Section \ref{GOE} and conclude that
the order-chaos transition cannot explain the decay-out.

Instead of Eq.~(\ref{2Aav}) let us now take the background decay amplitude  
to be\cite{Sargeant:2001aq}
\be\label{2Aavpoi}
\ov{A_{00}}=\f{\Ga_S}{E-E_0+i[\Ga_S+\ov{W}_{00}(E)]/2}.
\ee
Following \AA berg\cite{Aberg:1999a} 
we assume that $|0\ket$ only mixes with one ND doorway
state $|d\ket$ whose energy is $E_d$.
The state $|d\ket$ is subsequently mixed by the residual interaction 
with the remaining ND states, $|Q\ket$, $Q=1,...N$, 
having the same spin as $|0\ket$ and $|d\ket$. The self-energy is then
given by
\be\label{self}
\ov{W}_{00}(E)=|V_{0d}|^2\sum_{Q=0}^N\f{|c_d(Q)|^2}{E-E_Q+i\Ga_N/2},
\ee
where $V_{0d}$ is the interaction energy of $|0\ket$ and $|d\ket$ and
$c_d(Q)$ is the amplitude of $|d\ket$ in $|Q\ket$.
The $|Q\ket$ lie in the interval 
$L=ND$ where $D$ here denotes the mean spacing in energy of the $|Q\ket$.
 
Ignoring an energy shift of $|0\ket$, the 
background decay intensity becomes
\bea
I_\av=\f{\Ga_S}{2\pi}\int_{-\infty}^{\infty}dE
\f{1}{[E-E_0]^2+[\Ga_S+2\pi|V_{0d}|^2S_d(E)]^2/4},\hspace{4mm}
\label{Iavd}
\eea
where the doorway strength function is given by 
\be\label{strength}
S_d(E)=\f{\Ga_N}{2\pi}\sum_{Q=0}^N
\f{|c_d(Q)|^2}{(E-E_Q)^2+\Ga_N^2/4}.
\ee  
In Sargeant \emph{et al.}\cite{Sargeant:2001aq} the effect of the chaoticity 
of the ND states on $I_\av$ was investigated by varying the strength 
of the residual interaction of the $|Q\ket$ and their interaction with 
$|d\ket$, both being assumed to be proportional to the chaoticity 
parameter $\la$.
The limiting value $\la$=$0$ results in the $|Q\ket$ having
Poisson statistics (regularity) while $\la$=$1$ results 
in their having GOE statistics (chaos).
The value of $\la$ determines the shape of $S_d(E)$
which is precisely the strength function that was previously 
investigated as a function of $\la$\cite{Sargeant:1999qk}. 

Instead of studying the interpolation between the limits $\la$=$0$ and
$\la$=$1$ by numerically diagonalising random matrices and performing
ensemble averages as we did before\cite{Sargeant:1999qk,Sargeant:2001aq},
here we restrict ourselves to the limiting case $\la$=$0$. 
As $\la\rightarrow 0$, $c_d(Q)\rightarrow \delta_{dQ}$ so that
$S_d(E)$ reduces to the single Breit-Wigner term,
\be\label{BW}
S_d^{\rm{Poi}}(E)=\f{1}{2\pi}\f{\Ga_N}
{(E-E_d)^2+\Ga_N^2/4}.
\ee
When $\la\rightarrow 0$ \emph{and} $\Ga_N\rightarrow 0$, 
then $S_d(E)\rightarrow\delta(E-E_d)$. 
For non-zero $\la$, $S_d(E)$ broadens
with increasing $\la$ until when $\la=1$ it takes a form, 
independent of $\Ga_N$, which is well 
approximated by \cite{Sargeant:1999qk}
\be\label{SGOE}
S_d^{\rm{GOE}}(E)=\left\{
\begin{array}{cc}
1  ,&|E-E_d|\le L/2\\
0  ,&|E-E_d| >  L/2
\end{array}
\right..
\ee
Inserting Eq.~(\ref{SGOE}) into Eq.~(\ref{Iavd}) we find that 
$I_\av^{\rm{GOE}}=(1+\Ga^\da/\Ga_S)^{-1}$ as long as $\Ga_S+\Ga^\da\ll L$,
in agreement with Eq.~(\ref{2Iav}) of Section~\ref{GOE}. Gu \emph{et al.}
discuss the strength function further for $L$ finite\cite{Gu:2002qe} 
and $L$ infinite\cite{Gu:2002}. 
%%%%%%%%%%%%%%%%%%%%%%%%%%%%%%%%%%%%%%%%%%%%%%%%%%%%%%%%%%%%%%%%%%%%%%
\begin{figure}
\includegraphics[width=\textwidth]{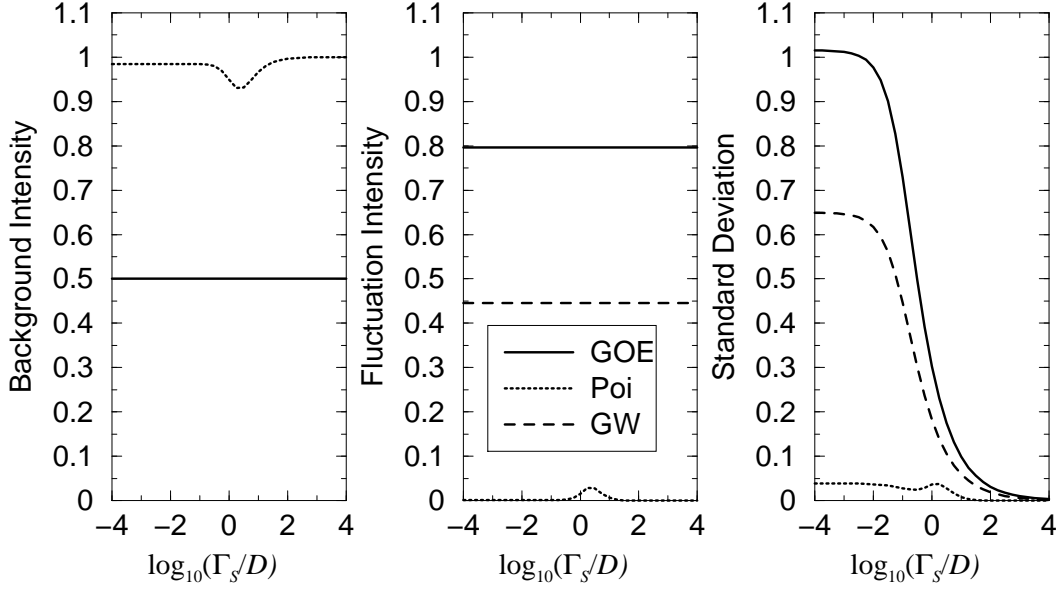}
\caption{\label{fig} The left graph shows the average background decay 
intensity versus $\log_{10}(\Ga_S/D)$ [The solid line was calculated using
Eq.~(\ref{2Iav}) and the dotted line using Eq.~(\ref{1Iavd})]. 
The middle graph shows the fluctuation contribution to the
average decay intensity versus $\log_{10}(\Ga_S/D)$
[The solid line was calculated using Eq.~(\ref{3ovIfl}) with
Eq.~(\ref{2Iav}) and the dotted line using 
Eq.~(\ref{3ovIfl}) with Eq.~(\ref{1Iavd}). The dashed line was calculated
using Eq.~(\ref{gufit})]. 
The right graph shows the standard deviation of the decay intensity 
(=$\sqrt{\ov{\left(\Delta I\right)^2}}$)
versus $\log_{10}(\Ga_S/D)$
[The solid line was calculated using Eq.~(\ref{3var}) with
Eqs.~(\ref{3ovIfl}) and (\ref{2Iav}) and the dotted line using 
Eq.~(\ref{3var}) with Eqs.~(\ref{3ovIfl}) and (\ref{1Iavd}). 
The dashed line was calculated using Eq.~(\ref{3var}) with Eq.~(\ref{gufit})]. 
The following numerical values were used to plot the graphs:
$\Ga^\da/\Ga_S=1$, $\Ga_N/D=0.1$ and $\De/D$=1.}
\end{figure}
%%%%%%%%%%%%%%%%%%%%%%%%%%%%%%%%%%%%%%%%%%%%%%%%%%%%%%%%%%%%%%%%%%%%%%
Eq.~(\ref{Iavd}) for $I_\av$ with $S_d=S_d^{\rm{Poi}}$ depends on 
four parameters: $\Ga_S$, $|V_{0d}|^2$, $\Ga_N$ and
the distance in energy separating $|d\ket$ from $|0\ket$, 
$\De=E_d-E_0$. It is useful to introduce a spreading width 
now defined to be $\Ga^\da=2\pi|V_{0d}|^2/D$.
Inserting Eq.~(\ref{BW}) into Eq.~(\ref{Iavd}) yields our result for $I_\av$
in the Poisson limit: upon making the change of integration variable 
$x=(E-E_0)/D$ this takes the form
\bea
I_\av^{\rm{Poi}}=\f{\Ga_S/D}{2\pi}\int_{-\infty}^{\infty}dx
\f{1}{x^2+\left(\Ga_S/D\right)^2\left[1+\f{\Ga^\da}{\Ga_S}\f{1}{2\pi}
\f{\Ga_N/D}{[x-\De/D]^2+(\Ga_N/D)^2/4}\right]^2/4}.
\label{1Iavd}
\eea
It is seen from Eq.~(\ref{1Iavd}) that, in addition to $\Ga^\da/\Ga_S$, 
$I_\av^{\rm{Poi}}$ is a function of a further three dimensionless variables:
$\Ga_N/D$, $\Ga_S/D$ and $\De/D$. 

The introduction of the ND doorway is analogous to including
another class of complexity in multistep compound preequilibrium
reactions\cite{Nishioka:1986,Friedman:1981}. 
Calculation of the fluctuation contribution to the average 
decay intensity and the variance in the presence of an ND doorway
is beyond the scope of this paper.
However, since Eq.~(\ref{3ovIfl}) for 
the fluctuation contribution and Eq.~(\ref{3var}) for the variance are
expressible in terms of the background contribution, $I_\av$, 
one can obtain
\emph{add hoc} estimates by simply substituting Eq.~(\ref{1Iavd}) 
into Eqs.~(\ref{3ovIfl}) and (\ref{3var}) (see Figure~\ref{fig}). 
A rigorous calculation of the
fluctuation contribution and variance when the ND states obey 
Poisson statistics may be
possible using the averaging method of Gorin\cite{Gorin:1999}.
 
The most important conclusion to be drawn from Figure \ref{fig} 
is that a measurement of the variance (standard deviation) of the
decay intensity determines the density of ND states $\rho=1/D$.
For comparison we have also calculated the standard deviation by inserting
the GW fit formula [Eq.~(\ref{gufit})] into our formula for the
variance [Eq.~(\ref{3var})].

Eq.~(\ref{1Iavd}) is formally identical 
to the branching ratio, $F_S$, calculated by 
Stafford and Barrett\cite{Stafford:1999gz} (SB).
They model the decay-out by allowing a single SD state
to mix with a single ND state. SB and 
Dzyublik and Utyuzh\cite{Dzyublik:2003} who develop the method of
SB further address the lack of knowledge of $\De$ by averaging
over $\De$ assuming a uniform distribution\cite{Vigezzi:1990a}. 
Cardamone \emph{et al.}\cite{Cardamone:2003} 
on the other hand assume a Wigner distribution, 
$P_{\rm{GOE}}(\De)=\pi\De/2\exp(-\pi\De^2/4)$. The preceding discussion
suggests that a Poisson distribution $P_{\rm{Poi}}(\De)=\exp(-\De)$ is
more appropriate.

An Addendum to Sargeant \emph{et al.}\cite{Sargeant:2001aq} 
is in preparation, where
we shall discuss the question of whether the order-chaos transition
can explain the decay-out when $\De$ is non-zero\cite{Aberg:2003}.

%\section*{Acknowledgements}
%We would like to thank ...........

%\appendix
%\section{First Appendix} %Empty argument \section{} yields `Appendix'. 
%
%\section{Second Appendix}

%\bibliography{sargeant,rmt,sd,compound,books}

\begin{thebibliography}{99}
%%%%%%%%%%%%%%%%%%%%%%%%%%%%%%%%%%%%%%%%%%%%%%%%%%%%%%%%%%%%%
% Some macros are available for the bibliography:
%  o for general use
%    \JL : general journals                 \andvol : Vol (Year) Page
%  o for individual journal 
%    \AJ   : Astrophys. J.           \NC         : Nuovo Cim.
%    \ANN  : Ann. of Phys.           \NPA, \NPB  : Nucl. Phys. [A,B]
%    \CMP  : Commun. Math. Phys.     \PLA, \PLB  : Phys. Lett. [A,B]
%    \IJMP : Int. J. Mod. Phys.      \PRA - \PRE : Phys. Rev. [A-E]     
%    \JHEP : J. High Energy Phys.    \PRL        : Phys. Rev. Lett.
%    \JMP  : J. Math. Phys.          \PRP        : Phys. Rep.
%    \JP   : J. of Phys.             \PTP        : Prog. Theor. Phys.     
%    \JPSJ : J. Phys. Soc. Jpn.      \PTPS       : Prog. Theor. Phys. Suppl.
% Usage:
%  \PRD{45,1990,345}          ==> Phys.~Rev.\ \textbf{D45} (1990), 345
%  \JL{Nature,418,2002,123}   ==> Nature \textbf{418} (2002), 123
%  \andvol{B123,1995,1020}    ==> \textbf{B123} (1995), 1020
%%%%%%%%%%%%%%%%%%%%%%%%%%%%%%%%%%%%%%%%%%%%%%%%%%%%%%%%%%%%%
  
\bibitem{Twin:1986}
P.~J. {Twin}, B.~M. {Nyak{\' o}}, A.~H. {Nelson} \emph{et~al.}, Phys. Rev.
  Lett. \textbf{57} (1986) 811.

\bibitem{Singh:2002}
B.~{Singh}, R.~{Zywina} and R.~B. {Firestone}, Nucl. Data Sheets \textbf{97}
  (2002) 241.

\bibitem{Nilsson:1995}
S.~G. Nilsson and I.~Ragnarsson, \emph{Shapes and Shells in Nuclear Structure}
  (Cambr., 1995).

\bibitem{Wadsworth:2002}
R.~{Wadsworth} and P.~J. {Nolan}, Rep. Prog. Phys. \textbf{65} (2002) 1079.

\bibitem{Lopezmartens:2003}
A.~{Lopez-Martens}, F.~Hannachi, A.~Korichi \emph{et~al.}, Acta Phys. Pol. B
  \textbf{34} (2003) 2195.

\bibitem{Lopezmartens:2000}
A.~{Lopez-Martens}, T.~{D{\o}ssing}, T.~L. {Khoo} \emph{et~al.}, Phys. Scr. 
\textbf{T88} (2000) 28.

\bibitem{Shimizu:1992}
Y.~R. {Shimizu}, F.~{Barranco}, R.~A. {Broglia} \emph{et~al.}, Phys. Lett. B
  \textbf{274} (1992) 253.

\bibitem{Yoshida:2001}
K.~{Yoshida}, M.~{Matsuo} and Y.~R. {Shimizu}, Nucl. Phys. \textbf{A696} (2001)
  85.

\bibitem{Lauritsen:2002}
T.~{Lauritsen}, M.~P. {Carpenter}, T.~{D{\o}ssing} \emph{et~al.}, Phys. Rev.
  Lett. \textbf{88} (2002) 042501.

\bibitem{Wilson:2003}
A.~N. {Wilson}, G.~D. {Dracoulis}, A.~P. {Byrne} \emph{et~al.}, Phys. Rev.
  Lett. \textbf{90} (2003) 142501.

\bibitem{Paul:2001}
E.~S. {Paul}, S.~A. {Forbes}, J.~{Gizon} \emph{et~al.}, Nucl. Phys.
  \textbf{A690} (2001) 341.

\bibitem{Khoo:1993}
T.~L. {Khoo}, T.~{Lauritsen}, I.~{Ahmad} \emph{et~al.}, Nucl. Phys.
  \textbf{A557} (1993) 83.

\bibitem{Krucken:2001we}
R.~Krucken, A.~Dewald, P.~von Brentano \emph{et~al.}, Phys. Rev. C \textbf{64}
  (2001) 064316.

\bibitem{Gu:1999bv}
J.-z. Gu and H.~A. Weidenm{\"u}ller, Nucl. Phys. \textbf{A660} (1999) 197.

\bibitem{Sargeant:2002sv}
A.~J. Sargeant, M.~S. Hussein, M.~P. Pato \emph{et~al.}, Phys. Rev. C
  \textbf{66} (2002) 064301.

\bibitem{Andreoiu:2003}
C.~Andreoiu, T.~D{\o}ssing, C.~Fahlander \emph{et~al.}, Phys. Rev. Lett.
  \textbf{91} (2003) 232502.

\bibitem{Sargeant:2001aq}
A.~J. Sargeant, M.~S. Hussein, M.~P. Pato \emph{et~al.}, Phys. Rev. C
  \textbf{65} (2002) 024302.

\bibitem{Aberg:1999a}
S.~{{\AA}berg}, Phys. Rev. Lett. \textbf{82} (1999) 299.

\bibitem{Kawai:1973}
M.~Kawai, A.~K. Kerman and K.~W. McVoy, Ann. Phys. \textbf{75} (1973) 156.

\bibitem{Ericson:1963}
T.~Ericson, Ann. Phys. \textbf{23} (1963) 390.

\bibitem{Ericson:1966}
T.~Ericson and T.~Mayer-Kuckuk, Annu. Rev. Nucl. Sci. \textbf{16} (1966) 183.

\bibitem{Bonche:1990}
P.~{Bonche}, J.~{Dobaczewski}, H.~{Flocard} \emph{et~al.}, Nucl. Phys.
  \textbf{A519} (1990) 509.

\bibitem{Sargeant:1999qk}
A.~J. Sargeant, M.~S. Hussein, M.~P. Pato \emph{et~al.}, Phys. Rev. C
  \textbf{61} (2000) 011302.

\bibitem{Gu:2002qe}
J.-z. Gu, L.~Gao and B.~Hu, Phys. Rev. C \textbf{66} (2002) 054312.

\bibitem{Gu:2002}
J.-z. {Gu}, L.~{Gao} and B.~{Hu}, Phys. Rev. E \textbf{66} (2002) 026208.

\bibitem{Nishioka:1986}
H.~Nishioka, J.~J.~M. Verbaarschot and H.~A. Weidenm\"uller, Ann. Phys.
  \textbf{172} (1986) 67.

\bibitem{Friedman:1981}
W.~A. Friedman, M.~S. Hussein, K.~W. McVoy \emph{et~al.}, Phys. Rep \textbf{77}
  (1981) 47.

\bibitem{Gorin:1999}
T.~{Gorin}, J. Phys. A \textbf{32} (1999) 2315.

\bibitem{Stafford:1999gz}
C.~A. Stafford and B.~R. Barrett, Phys. Rev. \textbf{C60} (1999) 051305.

\bibitem{Dzyublik:2003}
A.~Y. {Dzyublik} and V.~V. {Utyuzh}, Phys. Rev. C \textbf{68} (2003) 24311.

\bibitem{Vigezzi:1990a}
E.~{Vigezzi}, R.~A. {Broglia} and T.~{D{\o}ssing}, Nucl. Phys. \textbf{A520}
  (1990) 179c.

\bibitem{Cardamone:2003}
D.~M. {Cardamone}, C.~A. {Stafford} and B.~R. {Barrett}, Phys. Rev. Lett.
  \textbf{91} (2003) 102502.

\bibitem{Aberg:2003}
S.~{{\AA}berg}, Phys. Rev. C \textbf{68} (2003) 069801.

\end{thebibliography}
% Produces the bibliography via BibTeX.
%\bibliographystyle{npa}

\end{document}